# Per-bucket concurrent rehashing algorithms

Anton A. Malakhov

Intel Corporation
Anton.Malakhov@intel.com

**Abstract**

This paper describes a generic algorithm for concurrent resizing and on-demand per-bucket rehashing for an extensible hash table. In contrast to known lock-based hash table algorithms, the proposed algorithm separates the resizing and rehashing stages so that they neither invalidate existing buckets nor block any concurrent operations. Instead, the rehashing work is deferred and split across subsequent operations with the table. The rehashing operation uses bucket-level synchronization only and therefore allows a race condition between lookup and moving operations running in different threads. Instead of using explicit synchronization, the algorithm detects the race condition and restarts the lookup operation. In comparison with other lock-based algorithms, the proposed algorithm reduces high-level synchronization on the hot path, improving performance, concurrency, and scalability of the table. The response time of the operations is also more predictable. The algorithm is compatible with cache friendly data layouts for buckets and does not depend on any memory reclamation techniques thus potentially achieving additional performance gain with corresponding implementations.

*Categories and Subject Descriptors*: D.1.3 [**Programming Techniques**]: Concurrent Programming – *Parallel programming*; D.4.1 [**Operating Systems**]: Process Management – *Synchronization; concurrency; multiprocessing, multiprogramming, multitasking*; E.2 [**Data Storage Representation**] – *Hash-table representations.*

## 1. Introduction

Synchronization of state and data is often an unavoidable part of communication in applications executing on multi-core processors. While synchronization is sometimes necessary, it should be minimized to avoid overheads that may limit scalability or increase response time.

One of the fundamental building blocks in many applications is a hash table [3]. A hash table is a container that associates keys with values and uses a hash function to calculate placement positions for these pairs in a main array. To ensure that concurrent access to a hash table does not limit scalability or increase response time of an application, the hash table implementation must be optimized to minimize synchronization time.

However, designing efficient concurrent hash tables is challenging. An application may not know in advance the amount of items to be stored and therefore generic hash tables must be able to dynamically grow the underlying storage as items are added. In most implementations, growth of the storage requires full rehashing of the table, which limits access to the container until the rehashing is complete.

This work proposes a generic way to extend a hash table concurrently without a global synchronization, thus reducing the impact of table growth on scalability.

We implemented one of the variations of the algorithm and compared it against two implementations that represent two types of other known approaches for concurrent resizing of the hash table. Our implementation shows better performance and scalability on average.

### 1.1 Concurrent resizing problem

Hash tables are based on an array of *buckets*. Each bucket can contain or point to one or more key-value pairs [3]. Each lookup operation for a given *Key* uses a hash function $H$ to compute an index $i$ in the array of size $S$:

$$i = H(\text{Key}) \bmod S, \quad [F1]$$

where the value of $H$ is invariant for a given *Key*. It is important to align the size $S$ with number of pairs $n$ contained in the table, because when $n$ is much bigger than S, searching in a bucket takes more time as more pairs share the same bucket.

As it follows from [F1], the growth of a hash table invalidates the positions of all the stored pairs because these positions depend on the size of the array. The pairs need to be moved to new places, and this operation is known as *rehashing*. For concurrent algorithms, rehashing is challenging since many items need to be moved at a time.

### 1.2 Related work

The evolution of concurrent extensible hash tables is well described by Shalev and Shavit in [6]. To sum up, the best known lock-based implementations from the Java™ Concurrency Package [4] and Intel® Threading Building Blocks 2.1 [2] use several high-level locks to protect different parts of the main array. Both use chained hashing [3], and each bucket holds a linked list of key-value pairs. Another lock-based algorithm with high-level locks, hopscotch hashing [5], uses the open-addressed hash scheme where each bucket can contain only one data pair.

High-level locks allow limited concurrency when the work can be done for different parts of the array at a time but operations may be blocked when accessing buckets under the same lock.

The lock-free extensible hash table by Shalev and Shavir [6] avoids the necessity in explicit rehashing by maintaining a specifically sorted linked list of all the pairs in the table so that resizing does not invalidate the list. It does not require high-level synchronization but requires a memory reclamation support to enable erase operation. Also, a linked list is not a cache-friendly data layout (as discussed in hopscotch hashing paper [5]).

Our rehashing approach can be applied even for cache-friendly open-addressed hash tables and does not rely on the memory reclamation support.

## 2. Per-bucket concurrent rehashing

When the table needs to be extended, the proposed algorithm allocates as many buckets as already exist, and keeps the old buckets, thus doubling the number of buckets (*2S*). Each new bucket is logically mapped onto an existing (*parent*) bucket (Figure 1). It means that during rehashing of bucket $i_p$, $0 \leq i_p < S$, its pairs can be either left in place or moved into the new bucket $i_n$, $S \leq i_n < 2S$, such that

$$i_n \bmod S = i_p \quad [F2]$$

Proof: For any given key and its hash value $h = H(\text{Key})$, the above equation is equivalent to

$$(h \bmod 2S) \bmod S = h \bmod S$$

The last equation is trivially proved assuming $h = 2Sk + x$, where $k \geq 0$ and $0 \leq x < 2S$.

Each new bucket is marked as new (or *non-rehashed*) before allocation is finished by the publication of the new space, and rehashing is not required to follow before or immediately after the publication. Thus, resizing does not impact concurrency because it neither invalidates nor locks existing buckets.

After the new capacity is published, any operation that encounters a new bucket is redirected to the parent bucket(s) and moves (*rehashes*) part of its content into the new bucket(s). After the new bucket is updated it is marked as *rehashed*. Thus, the rehashing work is deferred and split across subsequent operations within the table.

This algorithm resembles linear hashing developed by W. Litwin in 1980 [7] but all buckets are allocated at once and split on demand in any order.

### 2.1 Power-of-two hashing

It is practical to use power-of-two hashing similarly to linear hashing, so light-weight bit-wise operations can be used instead of the division. For $S = 2^{level}$.

$$i = H(K) \,\&\, (2^{level}-1) \quad [F3]$$

The index of the parent bucket for a given i can be calculated as:

$$i_p = i \,\&\, (2^{\lfloor \log_2 i \rfloor} - 1) \quad [F4]$$

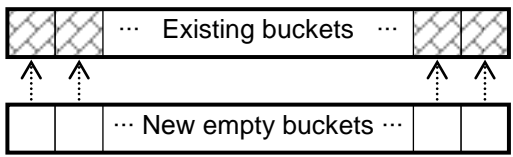

**Figure 1.** Allocation and mapping of the buckets.

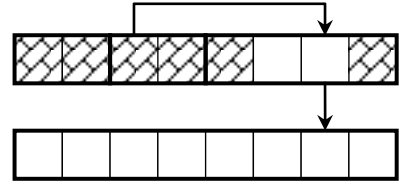

**Figure 2.** Recursive rehashing (two steps).

The subexpression $\lfloor \log_2 \rfloor$ gives the index of the most significant bit. So, this expression is actually a zeroing of this bit.

Using power-of-two hashing may hurt the even hash value distribution and uniform utilization of buckets. Thus, hash function *H* has to be more sophisticated. For example, Larson [8] uses a prime number to restore the distribution.

The first two buckets have no defined parent index, so they are *root* buckets. Otherwise, there is only one immediate parent for a bucket. And a parent bucket has one immediate child per each next level of capacity and even more indirectly.

And consequently, if a parent bucket is not rehashed until the next extension of the table or the capacity is increased by a few levels at once, the key-value pairs have to be split along more than one bucket.

### 2.2 Recursive rehashing

The simplest way to rehash a bucket is to access its parent and move necessary data between them. But if the parent is also "new", it must first be rehashed, leading to a recursive rehashing operation where a pair may be moved in one bucket and immediately into another. See Figure 2.

Besides excessive moves, another drawback of this scheme is that data is moved only when a new bucket is accessed, while accessing "old" buckets does not cause rehashing, which may slow down the search across such a bucket the next time.

### 2.3 Multilevel rehashing

The drawbacks of recursive hashing may be addressed by extending buckets with a more complex control field that indicates the number of segments rehashed. So, any operation accessing an "old" bucket can detect whether rehashing is necessary and move the data directly into all the new buckets available at once as shown in Figure 3.

Nevertheless, recursive rehashing may be a better choice when resizing is a rare event and there is no space for additional field in the bucket.

### 2.4 Generality

The proposed algorithm does not restrict implementers to a specific structure of buckets and corresponding implementations of the table operations. It depends only on a high-level scheme of allocations for the array and some simple flags in each bucket. So it may be used with various concurrent hash table algorithms which need to move the key-value pairs among the buckets while resizing.

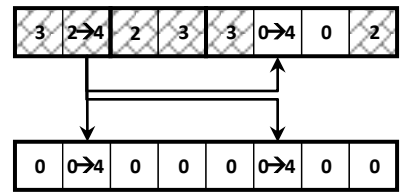

**Figure 3.** Multilevel rehashing (one turn).

### 2.5 Reactive synchronization

Any hash table that employs the proposed per-bucket rehashing algorithm must not only correctly synchronize the usual find, insert, and erase operations but also the per-bucket rehashing operations as well. However, bucket-level synchronization is not sufficient to maintain the correct global state of the table with per-bucket rehashing because the value of table capacity $S$ is always read before accessing a bucket (to calculate its index i) and can be changed right afterwards:

| Thread 1 | Threads 2, 3 |
|---|---|
| i = H( Key ) mod S | |
| | S = S*2 // table grow |
| | Rehash_bucket( i ) |
| | // Key can be moved from i |
| Lookup( Key, i ) | |

Therefore, if a key is not found during a search of the hash table, it is possible that it was moved by a concurrent rehashing of the bucket. To eliminate possibility of this race condition, explicit global-level synchronization is required but it hurts scalability and increases the hot path of every table operation.

Instead, the proposed algorithms handle this potential race condition using another approach. To make sure a key does not exist, the current capacity of the array is compared with the capacity used to compute the bucket index. If (1) they are not equal, (2) the bucket indexes computed with these values are different, and (3) the bucket with the old index was (or is being) rehashed towards new one, then the algorithm restarts the lookup operation.

For recursive rehashing, the last part is complicated because there is no direct indicator of whether the old bucket was used to rehash the new one. Instead, the next child bucket where the key can be found is checked for required rehashing state as shown in function is_race at Figure 5 and described in 3.1.

For multilevel rehashing, this check is trivial using information about rehashed segments stored in the bucket, which in fact is current capacity effective for the given bucket. The difficulty here is rather in algorithm of detecting a correct root parent while other threads can rehash it concurrently (not discussed in the paper).

Reactive synchronization is the key idea for lock-free synchronization of global consistency that is common for both rehashing algorithms. It does not impose the costs to the hot path and thus, it does not hurt concurrency while maintaining overall correctness.

### 2.6 Complexity

The complexities of the hash table operations depend on the buckets layout and implementation of the operations. In addition, concurrent rehashing requires restarting of lookup operations if the race condition is detected. But as shown below, the statistical probability of this event is negligibly small.

With recursive rehashing, deferred rehashing can lead to more cycles wasted for searching in an old bucket until accessing the related new bucket rehashes it. As we show below, the slowdown is noticeable when the insertion of unique keys takes the largest portion of all the operations within the table.

## 3. Performance

We implemented a simple chaining concurrent hash table with recursive rehashing in C++ using synchronization primitives from Intel's TBB library [2] like tbb::atomic and tbb::spin_rw_mutex. These are the same basic blocks that were used in tbb::concurrent_hash_map (as implemented in TBB 2.1) and tbb::concurrent_unordered_map (TBB 3.0) which we compare it against. The former uses high-level locks as we described above. And the latter is based on the original split-ordered list algorithm by Shalev and Shavir [6] and a well known extensible dynamic array algorithm [1] similar to the one implemented in tbb::concurrent_vector [2]. Our implementation uses the same approach to double the number of buckets. In addition, the structure of buckets is the same as the one in tbb::concurrent_hash_map.

All these similarities help to exclude the majority of factors that are not related to the concurrent rehashing itself. However, tbb::concurrent_unordered_map lacks a concurrent erase operation, which gives it an unfair advantage over the rivals due to a simplified synchronization protocol without memory reclamation support. Nevertheless, as we show in the performance evaluation section below, our implementation performs better in most cases.

### 3.1 Implementation details

Our implementation was carefully tested and released as part of a commercial product available also in open sources. Figure 5 represents the pseudo-code compiled from the real implementation and highlights the main flow related to concurrent resizing and recursive rehashing. However, for the sake of simplicity and readability of the code, we omit some details like a read-writer lock and the double-check idiom. Furthermore, this fragment does not include other operations like find and erase. They can be easily derived from the presented insert operation.

The first lines define the basic structure of the hash table. my_mask contains the total number of buckets minus one (S-1), and my_size contains the number of inserted pairs. It is important for the scalability to avoid false sharing by placing these variables in separate cache lines.

my_table is a *segment table* of a fixed size that stores pointers to arrays of buckets – *segments* as shown in Figure 4. As the table grows, new segments are added. Each added segment contains as many buckets as all of the preceding segments, thus doubling the total number of buckets at each growth of the table. The size of my_table is a number of bits in my_size.

The size of the first segments is two buckets. It leads to the simplest segment index computations formulas. At lines 14-16, segment_index_of calculates the index of the segment for the

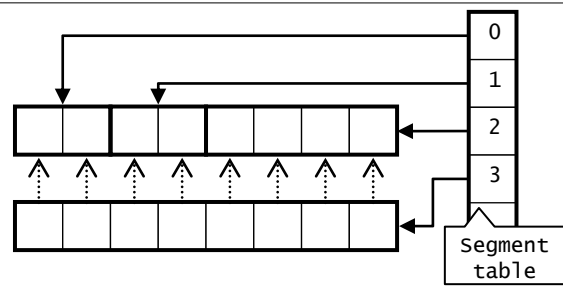

**Figure 4.** Segment table and mapping of a segment.

```
01 my_mask : atomic integer;
02 my_size : atomic integer;
03 my_table : array of pointers to buckets;
04
05 is_race ( h, m ) {
06   //find next applicable mask after old
07   while ( ( h & (m+1)) == 0 ) // test bit
08     m = m<<1;
09   m = m<<1;
10   //check whether it is [being] rehashed
11   return is_marked_new (h & m) == false;
12   // if true a lookup must be restarted
13 }
14 segment_index_of( i ) {
15   return Log2( index|1 );
16 }
17 // first global index of segment k
18 segment_base( k ) {
19   return (1<<k & inverse_bits(1));
20 }
21 acquire_bucket( i ) {
22   // find bucket i
23   k = segment_index_of( i );
24   j = i - segment_base( k );
25   b = my_table[k][j];
26   lock( b );
27   if( is_marked_new( b ) )
28     rehash_bucket( b, i );
29   return b;
30 }
31 rehash_bucket( b_new, i ) {
32   mark_rehashed(b_new);
33   // parent mask from the MSB of i
34   mask = ( 1<<Log2(i) ) - 1;
35   b_old = aquire_bucket( i & mask );
36   // get full mask for new bucket
37   mask = (mask<<1) | 1;
38   for each p in b_old {
39     h = H( p.key );
40     if( (h & mask) == i ) {
41       b_old.remove( p );
42       b_new.add( p );
43     }
44   }
45   unlock( b_old );
46 }
47 insert( Key ) {
48   k = 0;
49   h = H( Key );
50   m = my_mask.fetch();
51 label restart:
52   b = acquire_bucket( h & m );
53   p = b.search( key );
54   if( p == 0 ) {
55     m_now = my_mask.fetch();
56     if( (h & m)!=(h & m_now)
57       && is_race(h, m) ) {
58       unlock( b );
59       m = m_now;
60       goto restart;
61     }
62     // insert and set flag to grow
63     sz = my_size.fetch_and_increment();
64     b.insert( key );
65     // check load factor
66     if( sz >= m ) {
67       new_seg = segment_index_of( m+1 );
68       if( my_table[new_seg] == 0
69         && CAS(my_table[new_seg],marker,0))
70         k = new_seg; // processed below
71     }
72   }
73   unlock( b );
74   if( k ) { // new segment
75     // k > 0, first blocks pre-allocated
76     sz = 1 << k; // segment_size
77     segment = allocate_buckets( sz );
78     for each b in segment
79       mark_new( b );
80     my_table[k] = segment;
81     sz = sz * 2;// new value of capacity
82     my_mask.store( sz-1 ); // new mask
83   }
84 }
85
```

**Figure 5.** Pseudo code of chained concurrent hash table with recursive rehashing

bucket index i. Lines 18-20 compute the first bucket index in a given segment.

But using two buckets as the size of the first segment leads to excessive fragmentation of the cache lines if the first segments are allocated separately. In addition, resizing will happen more often, and contention on the first buckets will kill scalability.

Thus, we initially allocate enough (e.g. 512) buckets in a single block and just route segment pointers to appropriate positions inside the block. Therefore, a few pointers in the segment table will be set at once pointing inside a combined array of buckets. The logical bounds of the segments remain, but fragmentation is avoided. The first row of buckets in Figure 4 illustrates the joint allocation.

This allocation is done before any concurrent operation over the table which allows to place growth logic (lines 74-83) after insertion operation (lines 62-71). And thus, it helps to avoid unnecessary reading of the shared my_size counter before the actual insertion of a new key occurs. It is not known whether a key is unique or not until a lookup operation (lines 49-61, 73) finishes.

Each insert operation that adds a new pair in the hash table atomically increases the my_size counter and makes a decision whether to resize or not using its value as well as the current capacity (line 66).

The segment index is computed using the current capacity value (line 67) and if the pointer in the segment table is equal to zero then a compare-and-swap (CAS) operation sets the value to any non-zero value in order to select a thread that should perform the actual allocation of the new segment.

Before the publication of the new space, each bucket should be marked as new (lines 78-79). Publication ends by setting the current mask (i.e. capacity) to the new value (line 82).

After this point, any operation may access the new segment and the race condition described in section 2.5 can occur because lines 50 and 52 are not executed in one transaction.

If a key is not found (line 54), the current mask value is re-read and compared with the previous value applied on the hash code (line 56). Difference means that the concurrent growth affects the desired key, therefore rehashing status of an immediate child bucket should be checked (line 57).

The function **is_race** (lines 5-11) calculates the index of such a child by searching the next bit set after old mask and then checks the marker (line 11). It cannot just check a h&m_now bucket because a concurrent rehashing operation can move it to an intermediate bucket that stands between parent and the "target" bucket. For example, if m=3 and m_now = 15 (e.g. increased by two segments in a joint allocation), then for h = 14 the following situation is possible (see also Figure 2):

| Thread 1 | Other threads |
|---|---|
| acquire_bucket( 14 & 3 ) | |
| | my_mask = 15 |
| | rehash_bucket( 6 ); |
| b.search( key ) == 0 | // key moved to 6 |
| is_marked_new(14&15)==true | // race not detected |
| is_marked_new(14&7)==false | // race detected |

There is no partial rehashed state, so the rehash_bucket(6) cannot leave the pair in a parent bucket 2 even if the hash value and my_mask tell that they will be moved further to 14. Otherwise, the rehash_bucket(14) calling at line 35 acquire_bucket(6) will not initiate the rehashing of bucket 2 at line 28.

Instead, the rehash_bucket (lines 31-46) calculates a minimal mask for a specified new bucket (line 37) and compares hash values from the parent bucket cut by the mask with the index of the new bucket without respecting the current my_mask value.

The rehashing operation is initiated through the acquire_bucket (lines 21-30) called at line 52 under the lock (line 26) if the requested bucket is marked as new (line 27). Then, it calculates the index of the parent bucket at line 34 and calls the acquire_bucket again to lock the parent as well.

### 3.2 Performance evaluation

We used Intel® Threading Building Blocks (TBB) library [2] as a programming framework for evaluation of the described algorithm because our practical interest resided in a native programming using C++ without memory reclamation support. Also the library provides two implementations of concurrent hash table that represent both known types of concurrent resizing: algorithms with high-level locks (tbb::concurrent_hash_map) and split-ordered list (tbb::concurrent_unordered_map).

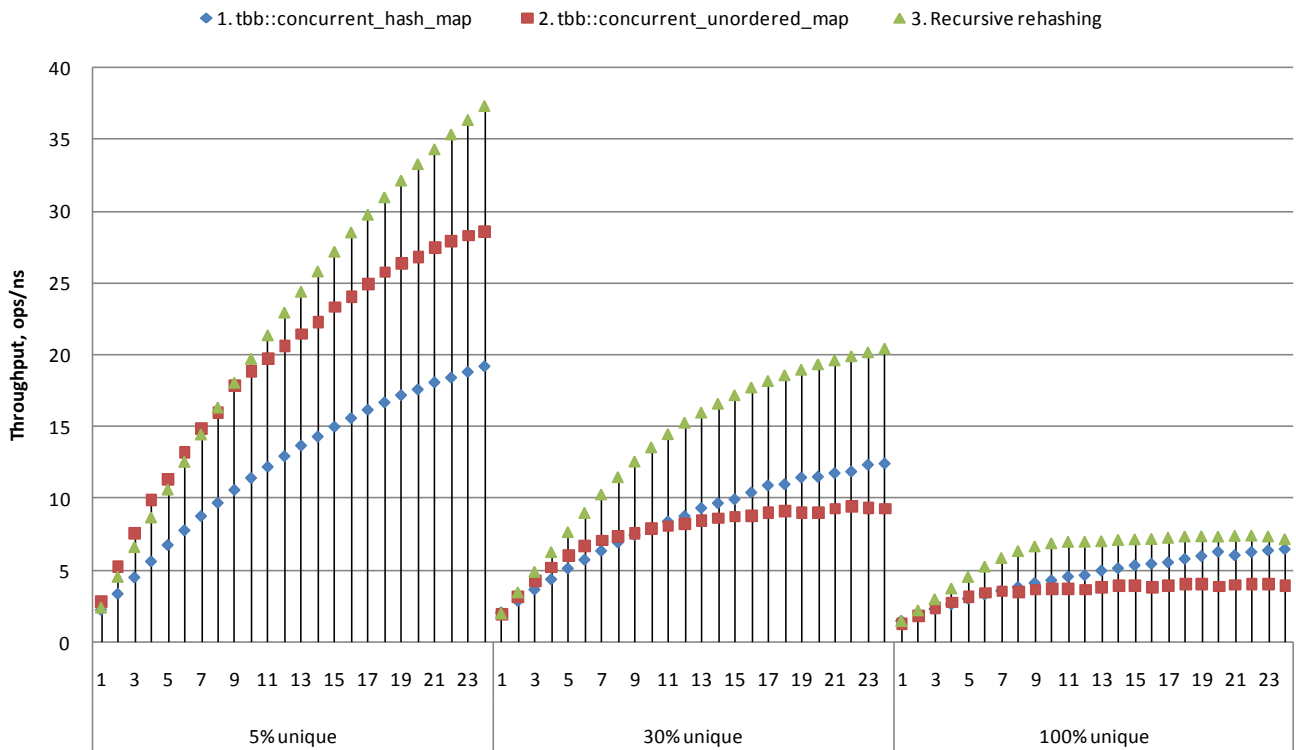

**Figure 6.** Throughput scalability of filling the table up to 2M pairs on various input rates

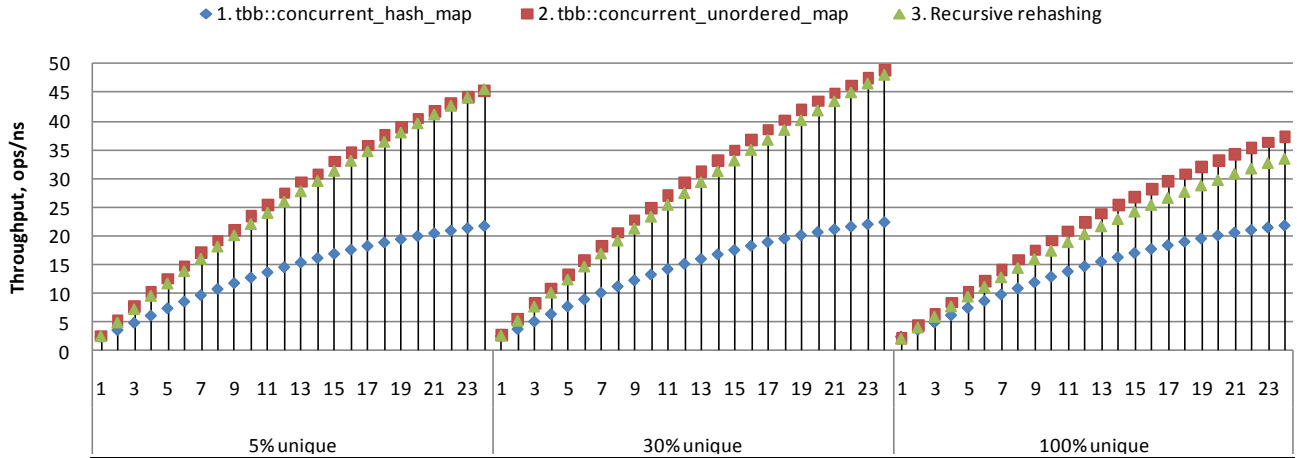
**Figure 7.** Throughput scalability of checking the table for existence of 2M inserted keys

We compared our hash table against these two on a four-socket machine with Intel® Xeon® Processors X7460 (6 cores per socket – 24 in total, 16M Cache, 2.66 GHz, 1066 MHz FSB) using a micro-benchmark specifically designed to highlight aspects of concurrent resizing of the hash table.

It was derived from a program that counts the number of some unique occurrences from the input data. We simplified the original task to exclude synchronization on the counters from the picture, so that it builds just a set of unique numbers from an input array. We filled the array evenly by using a pseudo-random number generator from the standard C library for various proportions of unique numbers. For example, for 5% of unique numbers, the same number is repeated 20 times on average. Together, it gives 5% of actual insertions and 95% are just lookups. However, in the beginning, there are more new keys occur than in the end.

In addition, a size of the source array correlates with input rates in order to produce the same number of unique keys and so exclude cache effects from the equation. Figure 6 shows the average throughput (axis Y) among the threads number (axis X) and input rates.

You can see that in most cases our implementation outperforms the rivals and scales better than split-ordered list for any rate. It also scales better than tbb::concurrent_hash_map for 5% but saturates after 8 threads for case when almost every input number is unique (shown as 100% but actually is about 99.95%).

As noted above, concurrent rehashing algorithm rehashes buckets on-demand. In addition, recursive rehashing does not initiates rehashing when accessing "old" buckets. It explains why the subsequent search over the table using the same input array (Figure 7) shows that recursive rehashing algorithm on "100%" scales worse than concurrent_unordered_map (which also defers bucket initialization). For other rates, these two implementations are almost the equal. However, as explained above, this simplified version of split-ordered list has the unfair advantage because does not synchronize lookup and erase operations.

The following statistics shows how more intensive insertions lead to less number of rehashed buckets and worse distribution of the key-data pairs along the hash table at the end of all operations.

| | Average, number of buckets | | |
|---|---|---|---|
| Unique,% | Rehashed | Empty | Overpopulated |
| 5 | 204850.0 | 48580.0 | 0.0 |

| | Average, number of buckets | | |
|---|---|---|---|
| Unique,% | Rehashed | Empty | Overpopulated |
| 10 | 2048482.6 | 48625.4 | 52.8 |
| 20 | 2032712.9 | 54454.9 | 8314.8 |
| 30 | 1963778.5 | 77963.0 | 43086.9 |
| 100 | 1245752.3 | 187169.7 | 591866.6 |

Here, the number of concurrent growths was 12.

The Figure 8 shows how the number of operation restarts grows along the number of threads, but in average it remains very small in comparison to the total number of operations. Thanks to that, reactive synchronization involves less overhead than explicit high-level synchronization.

## 4. Future work

Though this paper does not develop multilevel rehashing in details and does not present its evaluation, this version of the rehashing algorithm is expected to outperform recursive rehashing because it initiates rehashing faster for old buckets, acquires less locks, and moves key-value pairs directly. Also, it will feature a simpler race-detection code since there will be no intermediate rehashing of buckets. However, all these points relate rather to the growth condition and will be visible only for the case of the intensive

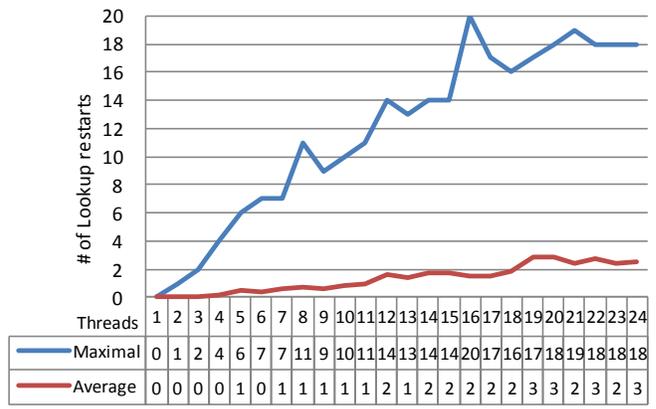
**Figure 8.** Number of restarts due to race condition

insertion of unique keys.

We may expect better scalability of an extensible hash table based on hopscotch hashing [5] combined with per-bucket concurrent rehashing.

Other directions of the research are lock-free underlying implementations, partial and cooperative variations of the per-bucket rehashing algorithm. They might be investigated to allow concurrent rehashing of the same buckets by different threads.

## 5. Conclusion

Our article presented a novel and generic approach to grow hash tables concurrently. It avoids high-level locks and does not depend on the memory reclamation techniques required by some related algorithms. Unlike related work, this algorithm detects races and restarts operations as needed.

It may be used in a wide range of hash table implementations of different types. We hope it will be inspiring to authors of the hash table implementations, as well as for researchers, because further research is necessary to find the best performing variant.

## Acknowledgments

Thanks to Mike Voss, Alexey Kukanov, Anton Eremeev, Volker Prott, and anonymous reviewers for valuable comments for this paper, as well as to other teammates who ever helped me with my work due to the algorithm.